\documentclass{article}
\usepackage{multirow}
\usepackage{spconf,amsmath,graphicx,amssymb}

\title{The CORAL++ algorithm for unsupervised domain adaptation of speaker recognition}
\name{Rongjin Li, Weibin Zhang, Dongpeng Chen}
\address{VoiceAI Technologies, Co. Ltd. Shenzhen, China}
\begin{document}
\ninept
\maketitle
\begin{abstract}
    State-of-the-art speaker recognition systems are trained with a large amount of human-labeled training data set. Such a training set is usually composed of various data sources to enhance the modeling capability of models. However, in practical deployment, unseen condition is almost inevitable. Domain mismatch is a common problem in real-life applications due to the statistical difference between the training and testing data sets. To alleviate the degradation caused by domain mismatch, we propose a new feature-based unsupervised domain adaptation algorithm. The algorithm we propose is a further optimization based on the well-known CORrelation ALignment (CORAL), so we call it CORAL++. On the NIST 2019 Speaker Recognition Evaluation (SRE19), we use SRE18 CTS set as the development set to verify the effectiveness of CORAL++. With the typical x-vector/PLDA setup, the CORAL++ outperforms the CORAL by 9.40\% relatively on EER.
\end{abstract}
\section{Introduction}
\label{sec:intro}

Speaker recognition is the task of recognizing a person's identity based on his or her voice \cite{reynolds2002overview}. In recent years, speaker recognition systems based on deep neural networks have achieved state-of-the-art performance in the community. Among them, the embedding-based methods can transform variable-length speech segments into fixed-dimensional vectors for scoring \cite{variani2014deep,snyder2017deep}. Neural networks are able to discriminate tiny differences among different speakers by learning from an extensive collection of labeled training set. Various training data provide rich nuisance factors for the model, making it considerably more robust in complex environments.

However, when a speaker recognition system is deployed in real world, it has to face the cross-domain problem where the domains in which the system is deployed differ from that it was trained. And during testing, the speech of enrollment and verification may be collected from different domains that were not presented during training. In this case, the testing data is called `in-domain' (inD) data while the training data is called `out-of-domain' (ooD) data. Unexpected cross-domain problems, such as cross-channels, cross-lingual, cross-devices, cross-codecs, duration shift and time-drifting, damage the performance of conventional algorithms. Since it is impossible to enumerate all cross-domain situations and collect all corresponding labeled data into the training set, speaker recognition systems suffer from performance degradation when encountering new challenges. Furthermore, labelling in-domain data for system re-training is expensive and time-consuming. Therefore, this issue has attracted lots of attention from both the academic and the industry \cite{sun2016return,alam2018speaker}. Many experts and researchers focus on closing the gap between inD data and ooD data by using unlabeled inD data sets.

To compensate for the performance degradation, many unsupervised domain adaptation strategies for back-end modeling have been proposed. For the x-vector (or i-vector) /PLDA pipeline \cite{snyder2018x,dehak2010front,prince2007probabilistic,ioffe2006probabilistic}, there are several main directions. For the model-based adaptation approaches, researchers aim to adapt the hyper-parameters of back-end models. In \cite{garcia2014unsupervised}, the authors proposed to adapt the between-class and within-class covariance matrices of PLDA model. Kong Aik \emph{et al.} proposed CORAL+ to compute the pseudo in-domain within and between class covariance matrices to regularize the corresponding matrices of PLDA \cite{lee2019coral+}. Secondly, the feature-based adaptation algorithm is simple and effective without introducing new models, essentially providing more salient features for subsequent model-based methods. The CORAL algorithm is proposed to align the covariance between out-of-domain and in-domain embeddings via a whitening and re-coloring process \cite{sun2016return, alam2018speaker}. In \cite{bousquet2019robustness}, a feature-Distribution Adaptor (fDA) is proposed to avoid the influence of residual components and inaccurate information during adaptation. In addition, there are other methods to eliminate the domain-mismatch problem, such as neural network fine-tuning, metric learning loss functions in networks \cite{duroselle2020metric} and scoring framework \cite{li2020principle}.

In this work, we present an optimized CORrelation ALignment (CORAL) algorithm that works directly on raw embeddings. Hence, we refer to the new algorithm as CORAL++. CORAL++ focuses on aligning the second-order statistics, i.e., the covariance matrices, through a controllable regularization of residual components and a flooring constraint of normalized eigenvalue-spectrum. The raw covariance matrix estimated from sparse in-domain data is usually unreliable and contains many various nuisance factors in different in-domain data. It is impractical to directly use such raw covariance matrix to make adaptation, so our proposed algorithm makes the estimated in-domain covariance matrix more robust based on the assumption that larger eigenvalues (variances) are crucial to the target domain while small eigenvalues are unreliable. \cite{bousquet2019robustness} proposed to filter out these worthless eigenvalues through a constant threshold. But it is difficult to find a suitable threshold for different data sets, especially when the in-domain data are gathered from multiple data sources. In order to emphasize the important eigenvalues more effectively and stably, we propose to use Z-score normalization on the eigenvalue spectrum and then apply a flooring constraint to remove those nuisance components. Then the in-domain covariance matrix is reconstructed to recolor the embeddings extracted from the out-of-domain data. Finally, the recolored embeddings are used to train the back-end models, e.g., PLDA. We carried out experiments on the NIST SRE19 CTS Challenge, and the corresponding development set includes the SRE18 CTS Dev and Eval sets \cite{sadjadi20202019}.

The rest of this paper is organized as follows. Section 2 reports the theory of CORAL and fDA. Both are relevant to our work. In section 3, we discuss the details of domain adaptation and CORAL++. The experimental setup is presented in Section 4 while the results are listed in Section 5. Finally, we conclude in Section 6.

\section{Related Work}
\label{sec:format}

The typical setting of unsupervised domain adaptation is that there is a universal model and a small amount of unlabeled in-domain data. For the x-vector / PLDA pipeline paradigm, it is simple and effective to adopt a feature-based adaptive strategy on training embeddings.

\subsection{Correlation Alignment}
\label{sec:format}

Addressing the domain-mismatch problem is critical for computer vision and speaker recognition. The main idea of CORrelation ALignment (CORAL) algorithm is to minimize the distance between the covariance of out-of-domain and in-domain embeddings \cite{sun2016return,alam2018speaker}. Suppose $D_I$ and $D_O$ are the $D$-dimensional embeddings of inD and ooD data sets respectively, i.e., $D_I \triangleq \{\vec{y_i}\}$, $\vec{y} \in \mathbb{R}^D$, and $D_O \triangleq \{\vec{x_i}\}$, $\vec{x} \in \mathbb{R}^D$. In addition, $C_I$ and $C_O$ are the covariance matrices of $D_I$ and $D_O$ respectively. The CORAL algorithm aims to find a transform matrix $A$ that minimize the Frobenius norm between the transformed out-of-domain covariance matrix and the in-domain covariance matrix, i.e.,

\begin{equation}\begin{split}\label{e1}
  A^* &= \mathop{\arg \min}_{A}\left \| C_{\widehat{O}} - C_I \right \|_{F}^{2} \\
&= \mathop{\arg \min}_{A}\left \| A^{\top} C_O A - C_I \right \|_{F}^{2}
\end{split}\end{equation}

\noindent where $C_{\widehat{O}}$ is the transformed ooD covariance matrix. There is an analytic solution for the above function \cite{sun2016return}.

\begin{equation}\begin{split}\label{e2}
  A^{*\top} = C_I^{\frac{1}{2}} C_O^{-\frac{1}{2}}
\end{split}\end{equation}

As can be seen, the optimal transformation matrix $A^*$ can be further decomposed into two parts: the first part $C_O^{-\frac{1}{2}}$ whitens the ooD data while the second part $C_I^{\frac{1}{2}}$ re-colors it with the inD covariance matrix. As also suggested in \cite{sun2016return}, in practice an identity matrix is usually added to the covariance matrix to make it full rank for the sake of efficiency and stability. Thus we can perform the classical whitening and coloring. This is advantageous since: 1) it is faster as singular value decomposition (SVD) on the original covariance matrix might slow to converge; 2) the process is more stable. That is

\begin{equation}\begin{split}\label{e2}
A^{*\top} = (C_I + \textbf{I})^{\frac{1}{2}}(C_O + \textbf{I})^{-\frac{1}{2}}
\end{split}\end{equation}

\noindent A pseudocode of CORAL algorithm is presented in Algorithm 1.

\begin{table}[h]
\begin{tabular}{l}
\hline
\textbf{Algorithm 1}: CORAL for Unsupervised Domain Adaptation \\ \hline
\textbf{Input}: out-of-domain data $D_O$, in-domain data $D_I$  \\
\textbf{Output}: Adapted out-of-domain data $D^*_O$      \\
$C_O = cov(D_O) + eye(size(D_O, 2)) $                  \\
$C_I = cov(D_I) + eye(size(D_I, 2)) $                  \\
$D_O = D_O * C_O^{-\frac{1}{2}}  $       \qquad \qquad \quad     $\%$ whitening out-of-domain data     \\
$D^*_O = D_O * C_I^{\frac{1}{2}} $       \qquad \qquad \quad     $\%$ re-coloring in-domain data       \\ \hline
\end{tabular}
\end{table}

\subsection{Feature-Distribution Adaptor}
\label{sec:format}

In paper \cite{bousquet2019robustness}, the authors proposed a method to deal with the problem that the in-domain covariance matrix usually is not reliable in speaker recognition. In a typical x-vector/PLDA setup, generally, the covariance matrix is $512 \times 512$ and only several thousand samples are available to train the covariance matrix. The authors argued that only large eigenvalues reflect the true characteristics of in-domain data, and small eigenvalues are noisy and unreliable. A flooring mechanism is thus proposed to keep large components in the in-domain covariance matrix for re-coloring. Moreover, the authors proposed to firstly apply by-domain mean adaptation to inD and ooD embeddings to eliminate mean-shift in cross-domain applications. The whole algorithm is shown in Algorithm 2. The $P$ and $\Delta$ are eigenvector and diagonal eigenvalues matrices, respectively.

\begin{table}[h]
\begin{tabular}{l}
\hline
\textbf{Algorithm 2}: feature-Distribution Adaptor \\ \hline
Apply by-domain mean adaptation to inD and ooD vectors. \\
Compute covariance matrices $C_I$, $C_O$ of inD and ooD data. \\
Compute SVD of $C_O^{-\frac{1}{2}}C_IC_O^{-\frac{1}{2}} = P\Delta P^{\top}$ \\
Compute matrix $\widehat{\Delta}$ such that $\widehat{\Delta}_{i,i} = max(1,\Delta_{i,i})$ \\
For each ooD vector $x$ do $x \leftarrow (C_O^{\frac{1}{2}}P\widehat{\Delta}^{\frac{1}{2}}P^{\top} C_O^{-\frac{1}{2}})x $ \\ \hline
\end{tabular}
\end{table}

For inD covariance, the feature-Distribution Adaptor computes eigenvalue-spectrum in the `whitened' space, which appears to be efficient to retain the specific information of target domain.

\section{The CORAL++ Algorithm}
\label{sec:pagestyle}

The object of CORAL is to minimize the matrix distance between $C_O$ and $C_I$, i.e., function (1). Analytically, the transformation matrix $A$ can be decomposed into two parts $C_I^{\frac{1}{2}}$ and $C_O^{-\frac{1}{2}}$, and only the $C_I^{\frac{1}{2}}$ of re-coloring process is decided by in-domain data. Obviously, the effect of re-coloring is so critical. And given limited amount of development sets, the estimation of $C_I^{\frac{1}{2}}$ is probably not reliable, so we need to further optimize $C_I^{\frac{1}{2}}$. The assumption of CORAL++ is that large values of eigenvalue spectrum are reliable in $C_I$, while those small ones are not reliable and need to be filtered out.

CORAL++ focuses on optimizing the eigenvalues of $C_I$ and then reconstructing $C_I$ using the optimized eigenvalues. Specifically, given the $D$-dimensional symmetric covariance matrix $C_I$, we compute the eigenvalues through eigenvalue decomposition. The raw eigenvalues varies largely and it is hard for us to set a universal threshold to filter out those unimportant ones. On the other hand, only the relative importance of eigenvalues is useful for the re-coloring process. Thus, we first propose to normalize the eigenvalues to have zero mean and unit variance through Z-score normalization. Suppose the eigenvalues are $s_i$ where $i=1,2, \cdots, D$. We have

\begin{equation}\begin{split}\label{e3}
  \widehat{s}_i = \frac{s_i - \mu_s}{\sigma_s}, i = 1, 2, \cdots, D
\end{split}\end{equation}

\noindent where $\mu_s$ and $\sigma_s$ are the mean and variance of eigenvalues respectively. After normalization, we can compare eigenvalues measured at different scales. It is easy to understand how good a certain eigenvalue is relative to the entire group. Then we set a universal threshold $\alpha$ to filter out those unreliable eigenvalues. That is

\begin{equation}\begin{split}\label{e4}
  v_i = max(\alpha, \widehat{s}_i), i = 1, 2, \cdots, D
\end{split}\end{equation}

\noindent where $\alpha$ is a variable used to determine the retention of eigenvalue components. Those elements with very low or even negative (after normalization) variances should be discarded by the $max(\cdot)$ operation. This step ensures that only those components with large variance are propagated to the transformation matrix $A$.

The traditional whitening is adding a small regularization parameter $\lambda$ to the diagonal elements of covariance matrix to explicitly make it full rank. The authors in \cite{sun2016return} argue that the performance of final system is insensitive to the value of $\lambda$ in computer vision and thus an identity matrix $\textbf{I}$ is used in \cite{sun2016return}. However, we found that the flexible adjusting the hyper-parameter $\lambda$ is helpful in domain-mismatch speaker recognition task. Adding a $\lambda$ that is too large will compress the relative differences between variances.

By combining all the points above, the proposed CORAL++ algorithm is shown in Algorithm 3. The $\alpha$ and $\lambda$ are the hyper-parameters and will be analyzed in Section 5.

\begin{table}[h]
\begin{tabular}{l}
\hline
\textbf{Algorithm 3}: CORAL++ for Unsupervised Domain Adaptation \\ \hline
\textbf{Input}: out-of-domain data $D_O$, in-domain data $D_I$  \\
\textbf{Output}: Adapted out-of-domain data $D^*_O$      \\
$C_O = cov(D_O) $                  \\
$C_I = cov(D_I) $                  \\
$EVD(C_I) \rightarrow P * diag(s) * P^{\top}$                 \quad \ \ \           $\%$ symmetric QR method \\
$\widehat{s}= Z(s)$                                           \qquad \qquad \qquad \qquad \qquad \quad \  \ $\%$ Z-score on eigenvalues \\
$v_i = max(\alpha, \widehat{s}_i)$                            \qquad \qquad \qquad   $\%$ applying a flooring constraint\\
$\widehat{C}_O = C_O + \lambda * eye(size(D_O, 2))$   \\
$\widehat{C}_I = P * diag(v) * P^{\top} + \lambda * eye(size(D_I, 2))$  \\
$D_O = D_O * \widehat{C}_O^{-\frac{1}{2}} $                   \qquad \qquad \quad   $\%$ whitening out-of-domain data     \\
$D_O^* = D_O * \widehat{C}_I^{\frac{1}{2}}$                   \qquad \qquad \quad   $\%$ re-coloring with in-domain data       \\ \hline
\end{tabular}
\end{table}

As can be seen from Figure 1, the usage of CORAL++ is the same as CORAL. Both are used as the first module of back-end system \cite{li2020voiceai}. The raw training embeddings are adapted by CORAL++ (or CORAL, or fDA) firstly, and then centering, principal component analysis (PCA), length normalization (LN), linear discriminant analysis (LDA) and Gaussian probability LDA (GPLDA) are successively trained by the adapted training embeddings. In addition, the CORAL+ algorithm is used to adapt the GPLDA model with the development embeddings and finally an adapted GPLDA is produced by interpolating the GPLDA with the CORAL+ model. The trials scores are then computed with the adapted GPLDA. To thoroughly compare different cross-domain adaptation techniques, we also used cosine distance scoring (CDS) in our experiments.

\begin{figure}[h]
  \centering
  \includegraphics[width=8cm]{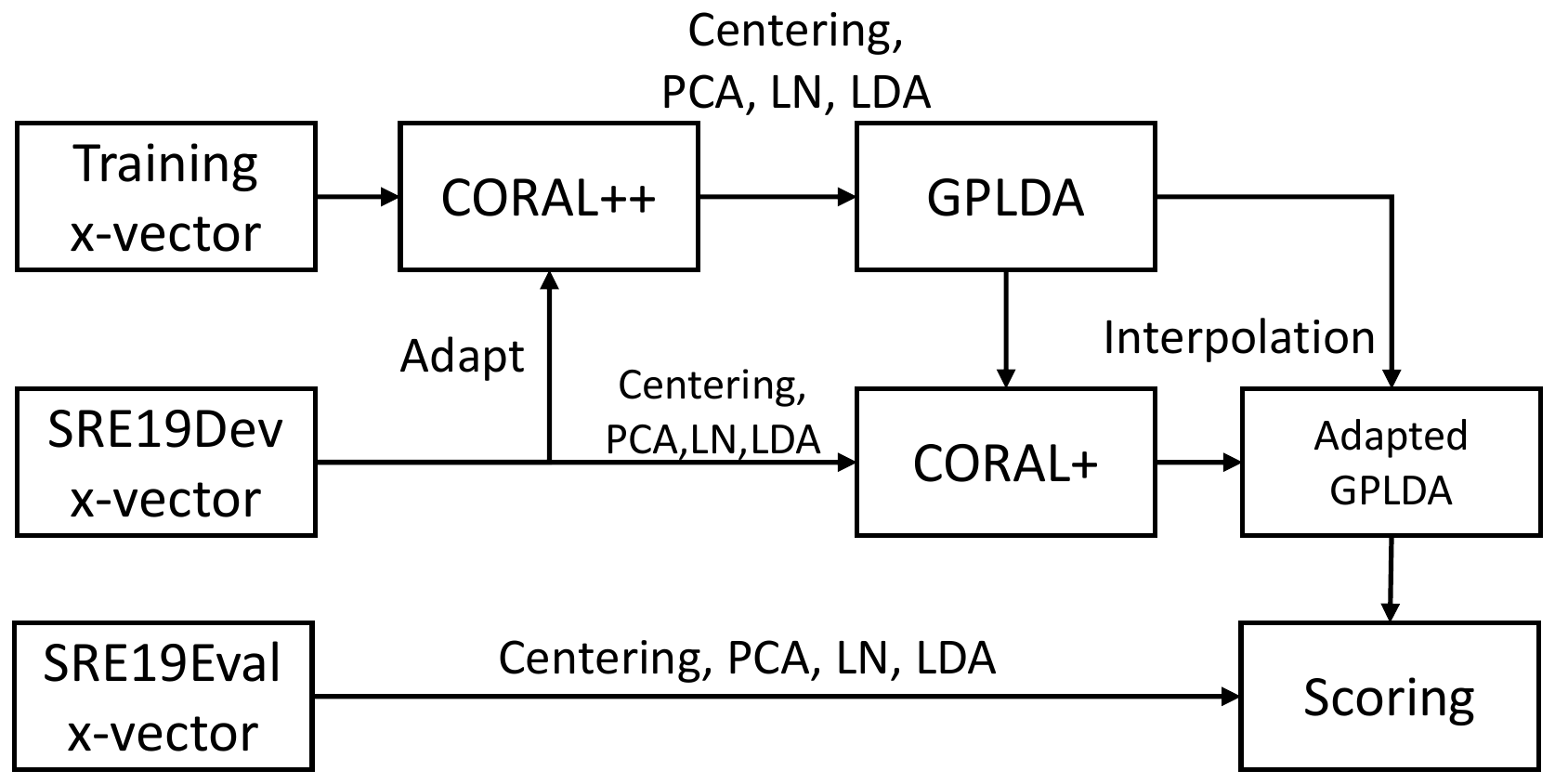}\\
  \caption{Flow Diagram of Back-end Optimized Strategies}\label{f1}
\end{figure}

\section{Experimental configuration}
\label{sec:typestyle}

We carried out our experiments with x-vectors of factorization time-delay neural networks (FTDNN) \cite{povey2018semi}. The effectiveness of CORAL++ is evaluated on the NIST SRE 2019 CTS Challenge.

\subsection{Training setup}

We used Switchboard (SWBD), NIST SREs, MIXER 6, Vox-Celeb, CN-Celeb \cite{fan2020cn} and Librispeech to train the neural networks. The SWBD corpora consist of SWBD 2 Phase 1, 2 and 3, and SWBD Cellular 1 and 2. The NIST SREs corpora consist of 2004 to 2010. For the MIXER6 corpus, we just used the telephone phone calls part. The Vox-Celeb and CN-Celeb corpora consist of 1 and 2, and we concatenated all segments of same utterance into a single one. The Librispeech corpora consists of train-clean and train-other parts. We also did data augmentation with additive noise and reverberation. The noise sets include MUSAN \cite{musan2015} and RIRS\_NOISES \cite{ko2017study}. Meanwhile, since encoding-decoding of speech is lossy during communication and storage, we used the MP3 codecs to simulate such a process. Finally, utterances that are shorter than 5 seconds and speakers with less than 8 utterances were all discarded.

All speech was down-sampled to 8KHz if the original recordings were 16KHz speech. The dimension of log Mel filter-banks (F-bank) was set to 64. All the features were extracted every 10ms with a 25ms shift window and the valid frequency was limited to 20-7600Hz. We applied the energy-based voiced activity detection (VAD) and cepstral mean-normalization (CMN) with a sliding window of up to 3 seconds on these acoustic features.

In the SRE19 CTS Challenge, the SRE18 data set, which consists of an unlabeled set (SRE18Dev) and labeled enroll-test sets (SRE18Eval), is used as the development set (named as SRE19Dev in Figure 1). We used the entire SRE19Dev as the in-domain data for adaptation and it contains 17,524 utterances. The SRE19Dev and SRE19Eval data are collected from 8KHz PSTN and VOIP and are spoken in Tunisian Arabic \cite{sadjadi20202019}. The training x-vectors extracted from the NIST SREs, Vox-Celeb 1 and CN-Celeb 1 datasets were used as the out-of-domain data to train back-end models. The out-of-domain data sets consist of multi-lingual (English, Chinese, etc.), multiple sampling rates (8KHz and 16KHz), multiple data sources (landline phone, web video, etc.) and so on. All the above make the domain mismatch problem in this task very challenging.

Meanwhile, in order to verify the consistency of hyper-parameters, we set two independent experimental groups where we just used SRE18Dev as the in-domain data to adapt the training x-vectors and it contains 4,073 utterances. And then we analyzed the corresponding results on the SRE18Eval.

\subsection{Models setup}

\begin{table*}[!ht]
\setcounter{table}{2}
\caption{Performance comparison of FTDNN x-vector systems with different domain adaptation methods on the NIST SRE19 CTS Challenge. The two evaluation metrics are EER(\%)/min-Cost. The ratio is the percentage of random subset of available data. The fDA is the feature-Distribution Adaptor method. $\lambda $ and $\alpha$ were set to 0.1 and 0.5 respectively for the CORAL++ algorithm.}\label{table1}
\centering
\begin{tabular}{ll|llll|llll}
\hline
    \multicolumn{2}{c|}{Scoring} & \multicolumn{4}{c|}{PLDA}                                     & \multicolumn{4}{c}{Cosine}                         \\ \hline
    \multicolumn{1}{c|}{Method}  & Ratio & \multicolumn{1}{c}{raw} & \multicolumn{1}{c}{CORAL} & \multicolumn{1}{c}{fDA} & \multicolumn{1}{c|}{CORAL++} & \multicolumn{1}{c}{raw} & \multicolumn{1}{c}{CORAL} & \multicolumn{1}{c}{fDA} & \multicolumn{1}{c}{CORAL++} \\ \hline
    \multicolumn{1}{c|}{Softmax} & 100\%                   & 6.47/0.453 & 6.47/0.454 & 7.01/0.486 & \textbf{5.73/0.421} & 7.66/0.494 & 7.64/0.493 & 8.56/0.548 & \textbf{6.24/0.433} \\ \hline
\multicolumn{1}{c|}{\multirow{3}{*}{AM-Softmax}} & 100\%   & 5.16/0.375 & 5.21/0.380 & 5.50/0.402 & \textbf{4.72/0.354} & 5.93/0.402 & 6.20/0.415 & 7.20/0.466 & \textbf{4.99/0.366} \\ \cline{2-10}
\multicolumn{1}{c|}{}            & 50\%                    & 5.20/0.377 & 5.24/0.381 & 5.48/0.404 & \textbf{4.75/0.355} & 6.01/0.404 & 6.16/0.414 & 7.28/0.471 & \textbf{4.98/0.364} \\ \cline{2-10}
\multicolumn{1}{c|}{}            & 10\%                    & 5.26/0.380 & 5.42/0.390 & 5.58/0.404 & \textbf{4.80/0.359} & 6.01/0.409 & 6.57/0.429 & 7.33/0.470 & \textbf{5.07/0.369} \\ \hline
\end{tabular}
\end{table*}

The FTDNN was trained by PyTorch \cite{paszke2019pytorch}, while acoustic features and back-end models were implemented with Kaldi \cite{povey2011kaldi}. The FTDNN we used is described in \cite{li2020voiceai}. It reduces parameters by factorizing weight matrices in a semi-orthogonal manner. Skip connections are introduced between low-rank interior layers, where prior layers are concatenated to form the input of current layer. As for the objective function, we used both Softmax and AM-Softmax loss functions. AM-Softmax was used to minimize intra-speaker variation and maximize inter-speaker discrepancy \cite{wang2018additive, li2019boundary}. The margin was set to 0.35 and the scale was 64 for the AM-Softmax.

All network models were trained using the SGD optimizer and the cyclical learning rate (CLR) strategy based on the triangular2 policy. The weight decay of SGD was $3e^{-4}$, and the max and min learning rates were set at $1e^{-2}$ and $1e^{-4}$, respectively. We trained the network models for 4 epochs with a batch-size of 128. Then the 512-dimensional x-vectors were extracted for scoring. The PCA and LDA reduced the dimension of x-vectors from 512 to 200, and from 200 to 100, respectively. There are no fine-tuning of network and no score normalization for all scoring results.

\section{Results}
\label{sec:majhead}

All systems were evaluated on the SRE18 and SRE19 evaluation set. Metrics used for performance measurement is equal error rate (EER) and minimum detection cost (min-Cost). We used the same parameters to calculate EER and min-Cost as in \cite{sadjadi20202019}.

CORAL++ includes two parameters, i.e., $\lambda$ and $\alpha$. We carried out experiments to see how the system performance is influenced by these hyper-parameters. There are some constraints on both $\lambda$ and $\alpha$. Firstly, $\lambda$ must be positive to ensure that both $C_I + \lambda \mathbf{I}$ and $C_O + \lambda \mathbf{I}$ are of full rank. Secondly, $\alpha$ must be non-negative since negative components of eigenvalue spectrum are not allowed.

In Table 1, we chose one of x-vector systems for analysis, i.e., the FTDNN-AMSoftmax/PLDA system. We set the flooring constraint $\alpha$ to be a constant 0 and varied $\lambda$. We could see that larger regularization parameter $\lambda$ results to better performance when $\lambda$ is smaller than 3.0.  $\lambda$ is added to the diagonal elements of covariance matrices to explicitly make them full rank, and it improve the generalization ability of model.

\begin{table}[h]
\setcounter{table}{0}
\caption{Sensitivity of Covariance Regularization Parameter $\lambda$ on the NIST SRE18 and SRE19 CTS Challenge. The metric is EER(\%). The $\alpha$ is fixed to 0.}\label{table2}
\begin{tabular}{l|lllllll}
\hline
$\lambda$       & 0.1  & 0.5  & 1.0  & 1.5  & 2.0   &2.5  & 3.0   \\ \hline
SRE18Eval       & 6.30 & 5.61 & 5.58 & \textbf{5.56}  & 5.58  & 5.63 & 5.66  \\ \hline
SRE19Eval       & 5.80 & 5.00 & 4.83 & 4.81  & \textbf{4.80}  & 4.83 & 4.86  \\ \hline
\end{tabular}
\end{table}

We continued to carry out experiments where $\lambda$ was fixed. The results are shown in Table 2. We found that the best value for $\alpha$ is much smaller than that of $\lambda$. $\alpha$ is used to filter out unreliable components from the eigenvalue spectrum. If $\alpha$ is too large, it would filter out some meaningful components. We recommend to set $\alpha=0.5$. We also found that after the proposed Z-score normalization, the eigenvalues follow the normal distribution and the adjustment of $\alpha$ becomes easier and more stable across of different datasets.

\begin{table}[h]
\caption{Sensitivity of Flooring constraint Parameter $\alpha$ on the NIST SRE18 and SRE19 CTS Challenge. The metric is EER(\%). The $\lambda$ is fixed to 0.1.}\label{table3}
\begin{tabular}{l|lllllll}
\hline
$\alpha$     & 0.1  & 0.5  & 1.0  & 1.5  & 2.0  & 2.5  & 3.0       \\ \hline
SRE18Eval    & 5.81 & \textbf{5.53}  & 5.56 & 5.71 & 5.85 & 5.98 & 6.06      \\ \hline
SRE19Eval    & 5.20 & \textbf{4.72}  & 4.73 & 4.81 & 4.88 & 4.97 & 5.06      \\ \hline
\end{tabular}
\end{table}

Following the above experiments, we chose $\lambda=0.1$ and $\alpha=0.5$ for the CORAL++ algorithm. In Table 3, we compared the x-vector systems with different domain adaptation algorithms on the SRE19 CTS Challenge. The ``raw'' method means that we did not use any adaptation method for the training embeddings in Figure 1. As can be seen, no matter the PLDA scoring or the cosine scoring are used, we can almost draw the same conclusion for the four adaptation techniques in comparison. The PLDA is superior to the cosine distance scoring in cross-domain problems for four different approaches. The PLDA can compensate for channel differences and it can be combined with other model-based adaptation algorithms to further enhance performance, such as the CORAL+. Therefore, we will focus on the PLDA scoring below. The performance of fDA is disappointing. We found that the fixed parameter flooring constraint (i.e., 1) in Algorithm 2 cannot effectively highlight the vital components. The AM-Softmax comparison group is much better than the Softmax group. We might draw a conclusion that increasing the inter-speaker variance and decreasing the intra-speaker variance help overcome the cross-domain problem.

As for the CORAL++, it achieves the best performance in different settings. When the Softmax is used, the CORAL++ is better than the CORAL by 11.44\% and 18.32\% relatively and respectively on EER when PLDA and CDS are used. When the AM-Softmax is used, the CORAL++ outperforms the ``raw'' method by 8.53\% and 15.85\% relatively and respectively on EER when PLDA and CDS are used. Furthermore, we randomly sampled 50\% and 10\% of SRE19Dev to compare the performance. For each percentage ratio, we randomly did three experiments and use the median of outputs. We found that only a subset of in-domain data is used can improve the performance by CORAL++, which proves to be stable.

\section{Conclusion}
\label{sec:print}

In this study, we focused on how to effectively align the second order statistics of in-domain and out-of-domain data through unsupervised adaptation on back-end models. The proposed CORAL++ algorithm is highly efficient and reliable to handle complex cross-domain speaker recognition tasks without requiring labeled data.

CORAL++ has been examined on the well-known NIST SRE19 CTS challenge and yields excellent results consistently. Suppressing the smaller eigenvalues of covariance matrix and highlighting the larger ones helps alleviate the cross-domain problem of sparse in-domain data. We believe that the core idea of CORAL++ could also provide hints for the improvement of neural network training and adaptation in future works.

\section{Acknowledgement}
\label{sec:print}

This work is partially supported by the key research and development program of Guangdong Province (2019B010154003), the fundamental research funds of IFS, China(2021JB019).

%\vfill
%\pagebreak

\bibliographystyle{IEEEbib}
\bibliography{strings,refs}

\end{document}